\documentclass[letterpaper, 10pt]{article}
\usepackage{times,amsmath,amsfonts,graphicx} % ,epsfig
\usepackage[normalem]{ulem}
\usepackage[buttonsize=10px]{animate}  %dvipdfmx,
\usepackage[tight]{subfigure}
\usepackage{color, multirow} %, soul
\usepackage{url}
\usepackage{flushend}
\usepackage[nohyperlinks,smaller,printonlyused]{acronym}

\newcommand{\ModifDF}[1]{{#1}} % en version finale
\acrodef{hc}[HC]{Hierarchical Clustering}
\acrodef{tf}[TF]{time-frequency}
\acrodef{tfr}[TFR]{time-frequency representation}
\acrodef{stft}[STFT]{Short-Time Fourier Transform}
\acrodef{wt}[WT]{Wavelet Transform}
\acrodef{cwt}[CWT]{Continuous Wavelet Transform}
\acrodef{snr}[SNR]{Signal-to-Noise Ratio}
\acrodef{rqf}[RQF]{Reconstruction Quality Factor}
\acrodef{if}[IF]{instantaneous frequency}
\acrodef{ife}[IFE]{instantaneous frequency estimator}
\acrodef{ifes}[IFEs]{instantaneous frequency estimators}
\acrodef{cr}[CR]{chirp rate}
\acrodef{cre}[CRE]{Chirp Rate Estimator}
\acrodef{cres}[CREs]{Chirp Rate Estimators}
\acrodef{mse}[MSE]{Mean Squared Error}
\acrodef{rqf}[RQF]{Reconstruction Quality Factor}
\acrodef{emd}[EMD]{Empirical Mode Decomposition}
\acrodef{ssa}[SSA]{Singular Spectrum Analysis}
\acrodef{imf}[IMF]{Intrinsic Mode Functions}
\acrodef{tftb}[TFTB]{Time-Frequency ToolBox}
\acrodef{astres}[ASTRES]{``Analysis, Synthesis, Transformation by Reassignment, EMD and Synchrosqueezing''}
\acrodef{svd}[SVD]{Singular Value Decomposition}

\renewcommand{\iint}[0]{\int\!\!\!\!\!\int}

\newcommand{\etal}[0]{\textit{et al.} }

\newcommand{\ie}[0]{\textit{i.e.} }

\newcommand{\tw}[0]{(t, \omega)}
\newcommand{\ttw}[0]{{\scriptstyle (t, \omega)}}

\newcommand{\fs}[0]{F_s}
\newcommand{\ee}{\,\ensuremath{\mathbf{e}}}

\renewcommand{\Re}[0]{\text{Re}}
\renewcommand{\Im}[0]{\text{Im}}

\newcommand{\TT}[0]{\mathcal{T}}
\newcommand{\DD}[0]{\mathcal{D}}

\begin{document}
\title{Second-order Time-Reassigned Synchrosqueezing Transform: Application to Draupner Wave Analysis} %desagregating

%\author{\IEEEauthorblockN{Dominique Fourer$^\star$ and }
  %corresponding author: 
%\and
\author{Dominique Fourer and Fran\c{c}ois Auger}

\maketitle
%\thanks{\input{acknowledgments.tex}}

%\let\thefootnote\relax\footnotetext{\input{acknowledgments}}

\begin{abstract}
This paper addresses the problem of efficiently jointly representing
a non-stationary multicomponent signal in time and frequency.
We introduce a novel enhancement of the time-reassigned
synchrosqueezing method designed to compute sharpened and reversible representations of
impulsive or strongly modulated signals.
After establishing theoretical relations of the new proposed method
with our previous results, we illustrate in numerical experiments
the improvement brought by our proposal when applied on both synthetic and real-world signals.
Our experiments deal with an analysis of the Draupner wave record for which we provide
pioneered time-frequency analysis results.
\end{abstract}

% \fi
%\IEEEpeerreviewmaketitle

\section{Introduction}

%% Objectifs / contexte
Time-frequency and time-scale analysis \cite{cbook, tfbook, HlawatschBook2008, flandrin2018} aim at developing efficient
and innovative methods to deal with non-stationary multicomponent signals.
Among the common approaches, the \ac{stft} and the \ac{cwt} \cite{wavelet} are the simplest linear transforms which 
have been intensively applied in various applications such as audio \cite{klapuri_f0est}, biomedical \cite{steeg07}, seismic or radar.

%% problematique
Unfortunately, these tools are limited by the Heisenberg-Gabor uncertainty principle.
As a consequence, the resulting representations are blurred with a poor energy concentration and require
a trade-off between the accuracy of the time or frequency localization.
Another approach, the reassignment method \cite{reassignment0,reassignment} was introduced 
as a mathematically elegant and efficient solution to improve the readability of a \ac{tfr}.
The inconvenience is that reassignment provides non-invertible \ac{tfr}s which limits its interest to analysis or modeling applications.

%% etat de l'art synchrosqueezing
More recently, synchrosqueezing \cite{synchrosqueezing, synch_reass_overview} was introduced
as a variant of the reassignment technique due to its capability to provide sharpen and reversible \ac{tfr}s.
This reconstruction capability make this method continuously gaining interest since it paves the way of an infinite number
of synchrosqueezing-based applications such as noise removal \cite{pham2018novel}, signal components extraction or separation \cite{synchrosqueezing,flandrin05,fourer2018,flandrin2018}.

Nowadays, efforts are made to efficiently compute the synchrosqueezed version of several linear transforms
such as \ac{stft}, \ac{cwt} or S-transform \cite{fourer2016,fourer2017c} and to improve the localization
of strongly modulated signals using enhanced instantaneous frequency estimators \cite{second_order_sst,fourer2017,pham2017high}.
To deal with impulses and strongly modulated signals, a new variant of the synchrosqueezing was introduced 
and called time-reassigned synchrosqueezing
method \cite{he2019time}. However, this method cannot efficiently deal with mixed-content signals containing both
impulsive and periodic components.

%% proposed solution
In the present paper, we propose to introduce a novel transform called the second-order horizontal synchrosqueezing
aiming to improve the energy localization and the readability of the time-reassigned synchrosqueezing while remaining reversible.
To this end, we use an enhanced group-delay estimator which can be mathematically related to our previous results \cite{fourer2017}.

%% paper organization
This paper is organized as follows. In Section \ref{sec:signalproperties}, 
the proper definitions of the considered transforms with their properties are presented.
In Section \ref{sec:synchrosqueezing}, we introduce our new second-order time-reassigned synchrosqueezing transform.
Section \ref{sec:results} presents numerical experiments involving both synthetic and 
real-world signals. Finally, future work directions are given in Section \ref{sec:conclusion}.

%% EOF

%--------------------------
\section{Time-reassigned synchrosqueezing in a nutshell} \label{sec:signalproperties}

\subsection{Definitions and properties}

We define the \ac{stft} of a signal $x$ as a function of time $t$ and frequency $\omega$
computed using a differentiable analysis window $h$ as:
%--------------------------
\begin{align}
F_x^h\tw 
 &= \displaystyle\int_{\mathbb{R}} x(\tau) h(t-\tau)^* \ee^{-j \omega \tau}\, \mathrm{d}\tau \label{eq:stft} %\\
%% &= \ee^{-j \omega t} \displaystyle\int_{\mathbb{R}} x(t-\tau) h(\tau)^* \ee^{j\omega \tau} \mathrm{d}\tau. \label{eq:stft_alt} % \ee^{-j\omega t} 
\end{align}
%--------------------------
where $j^2\!\!=\!\!-1$ is the imaginary unit and $z^*$ is the complex conjugate of $z$.
A \ac{tfr} also called spectrogram is defined as $|F_x^h\tw|^2$.
Thus, the marginalization over time of $F_x^h\tw$ leads to:  % as defined by Eq. \eqref{eq:stft}
%--------------------------
\begin{align}
 \displaystyle\int_{\mathbb{R}} F_x^h\tw \,dt  &= \iint_{\mathbb{R}^2} h(t-\tau)^* x(\tau) \ee^{-j\omega\tau} \,dt d\tau\\
					  &= \iint_{\mathbb{R}^2} h(u)^* x(\tau)\ee^{-j\omega\tau} du d\tau \\
					  &= \int_{\mathbb{R}} h(u)^* du \int_{\mathbb{R}} x(\tau) \ee^{-j\omega\tau}d\tau \\
					  &= F_{h}(0)^* F_x(\omega) \label{eq:margstft}
\end{align}
%--------------------------
with $F_x(\omega)\!=\!\int_\mathbb{R} x(t)\ee^{-j\omega t}\,dt$ the Fourier transform of signal $x$.
%(\ie $F_{h^*}(0)$ is the Fourier transform at frequency $0$ of function $h(t)^*$).
Now, from Eq. \eqref{eq:margstft} one can compute the Fourier Transform of $x$ as:
%--------------------------
\begin{equation}
 F_x(\omega) = \frac{1}{F_{h}(0)^*} \displaystyle\int_{\mathbb{R}} F_x^h\tw \,dt
\end{equation}
%--------------------------
and the following signal reconstruction formula can be deduced after applying the Fourier inversion formula:
%----------------------
\begin{equation}
 x(t) = \frac{1}{2\pi F_{h}(0)^*} \displaystyle\iint_{\mathbb{R}^2} F_x^h(\tau, \omega) \ee^{j \omega t} d\tau d\omega. \label{eq:reconstruction}
\end{equation}
%----------------------

\subsection{Reassignment}

To improve the readability of a \ac{tfr}, reassignment moves the signal energy according to: $\tw\!\mapsto\!(\hat{t}_x{\scriptscriptstyle \tw}, \hat{\omega}_x{\scriptscriptstyle \tw})$,
where $\hat{t}_x\tw$ is a group-delay estimator and $\hat{\omega}_x\tw$ is an instantaneous frequency estimator \cite{reassignment}.
Both time-frequency reassignment operators $\hat{t}$ and $\hat{\omega}$ can be computed as follows 
in the \ac{stft} case \cite{behera2016theoretical,fourer2017c}:
%---------------------------
\begin{alignat}{3}
 \hat{t}_x{\scriptstyle\tw}&= \Re\left(\tilde{t}_x{\scriptstyle\tw}\right), {\rm with}\quad
 \tilde{t}_x{\scriptstyle\tw} &=& t - \frac{F_x^{\TT h}{\scriptstyle\tw}}{F_x^h{\scriptstyle\tw}} \label{eq:timereassop}\\
 \hat{\omega}_x{\scriptstyle\tw}&= \Im\left(\tilde{\omega}_x{\scriptstyle\tw}\right), {\rm with}\quad
 \tilde{\omega}_x{\scriptstyle\tw} &=& j\omega + \frac{F_x^{\DD h}{\scriptstyle\tw}}{F_x^h{\scriptstyle\tw}} \label{eq:freqreassop}
\end{alignat}
%---------------------------
where $\TT h(t)\!=\!t h(t)$ and $\DD h(t)\!=\!\frac{d h}{d t}(t)$ are modified versions of the analysis window $h$.

Finally, a reassigned spectrogram can be computed as $\text{RF}^h_x\tw =$
%----------------------
\begin{equation}
 \displaystyle\iint_{\mathbb{R}^2} |F_x^h(\tau,\Omega)|^2 \delta\left(t-\hat{t}_x(\tau,\Omega)\right) \delta\left(\omega-\hat{\omega}_x(\tau,\Omega)\right) \mathrm{d}\tau \mathrm{d}\Omega.\label{eq:reass_stft}
\end{equation}
%----------------------

The resulting reassigned spectrogram $\text{RF}_x\tw$ is a sharpened but non-reversible \ac{tfr} due to the loss of the phase information.

\subsection{Time-reassigned synchrosqueezed STFT}

To overcome the problem of non reversibility, synchrosqueezing proposes to move the signal
transform instead of its energy, to preserve the phase information of the original transform.

Hence, time-reassigned synchrosqueezed \ac{stft} can be defined as \cite{he2019time}:
%----------------------
\begin{equation}
 S_x^h\tw = \displaystyle\int_{\mathbb{R}} F_x^h(\tau,\omega) \delta\left(t-\hat{t}_x(\tau,\omega)\right)\, d\tau \label{eq:trsst}
\end{equation}
%----------------------
where $\hat{t}_x\tw$ corresponds to the time reassignment operator which is classically computed using Eq. \eqref{eq:timereassop}.
% %----------------------
% \begin{equation}
%  \hat{t}_x\tw = t - \Re\left( \frac{F_x^{\TT h}{\scriptstyle\tw}}{F_x^h{\scriptstyle\tw}}\right)
% \end{equation}
% %----------------------
% with $\TT h(t) = t h(t)$.

The marginalization over time of the resulting transform leads to:
% \frac{1}{F_{h^*}(0)}
%----------------------
\begin{align}
 \displaystyle\int_{\mathbb{R}} S_x^h\tw dt &= \displaystyle\iint_{\mathbb{R}^2} F_x^h(\tau,\omega) \delta\left(t-\hat{t}_x(\tau,\omega)\right)\, dt d\tau \\
					    &= \displaystyle\int_{\mathbb{R}} F_x^h(\tau,\omega) d\tau = F_{h}(0)^* F_x(\omega). \label{eq:tsst}
\end{align}
%----------------------

Hence, an exact signal reconstruction from its synchrosqueezed STFT can be deduced from Eq. \eqref{eq:tsst} as:
%----------------------
\begin{equation}
 x(t) = \frac{1}{2\pi F_{h}(0)^*}\displaystyle\iint_{\mathbb{R}^2} S_x^h(\tau,\omega) \ee^{j \omega t} \, d\tau d\omega.
 \label{eq:reconsttsst}
\end{equation}
%----------------------

%% EOF
%---------------------------------------
\begin{figure*}[!ht]
 \subfigure[spectrogram]{\includegraphics[width=0.33\textwidth]{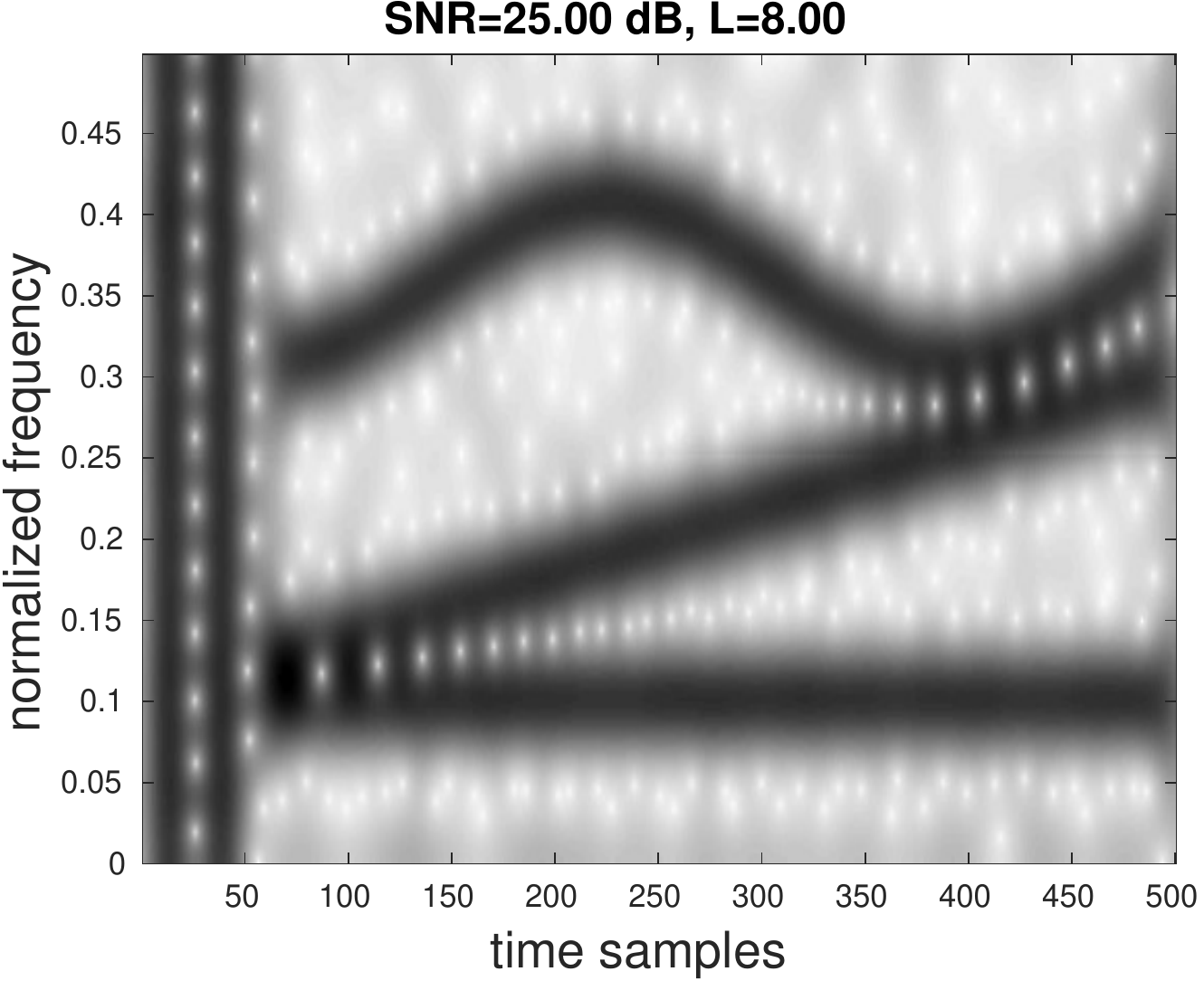}}~\subfigure[synchrosqueezing]{\includegraphics[width=0.33\textwidth]{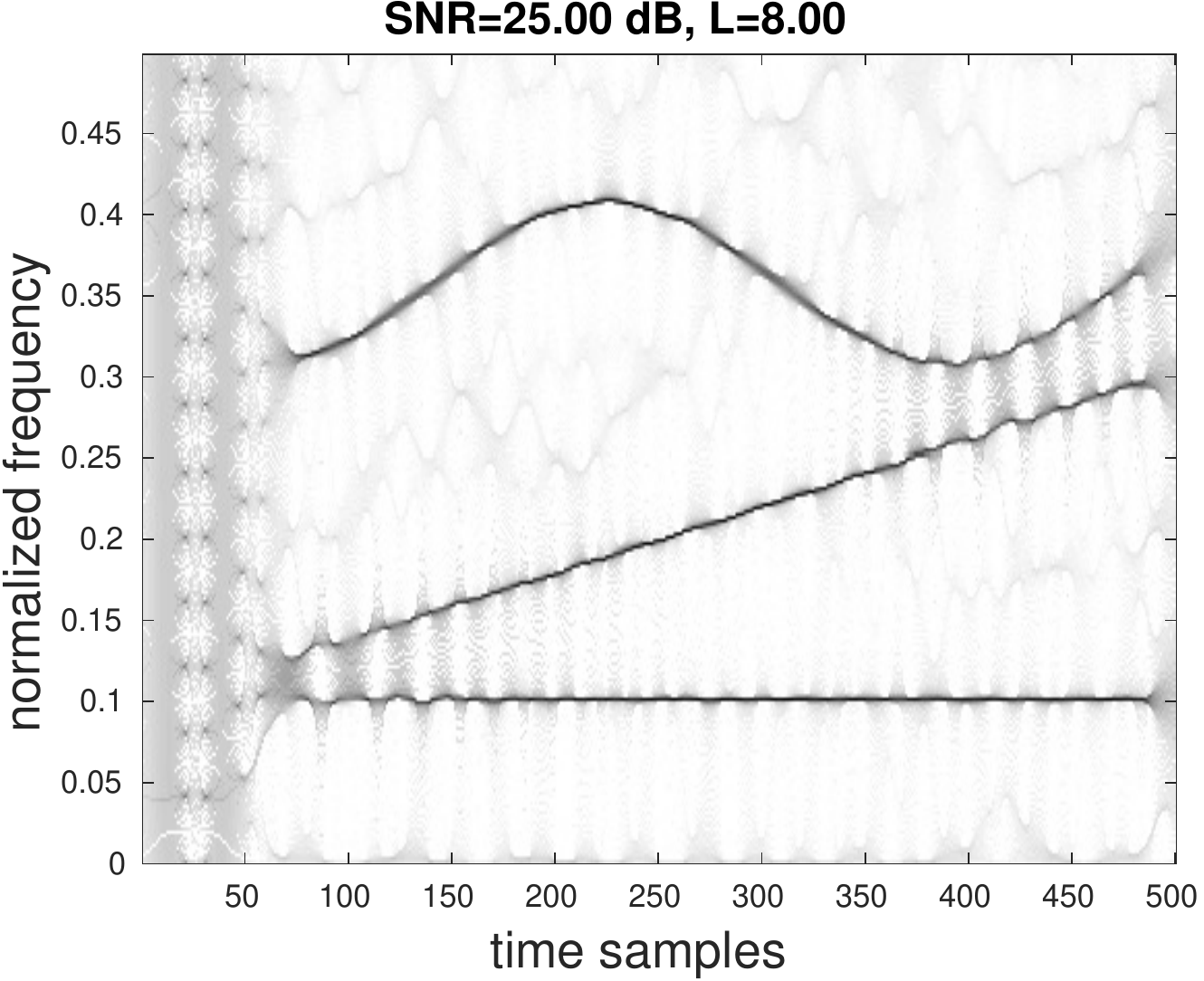}}~\subfigure[second-order vertical synchrosqueezing]{\includegraphics[width=0.33\textwidth]{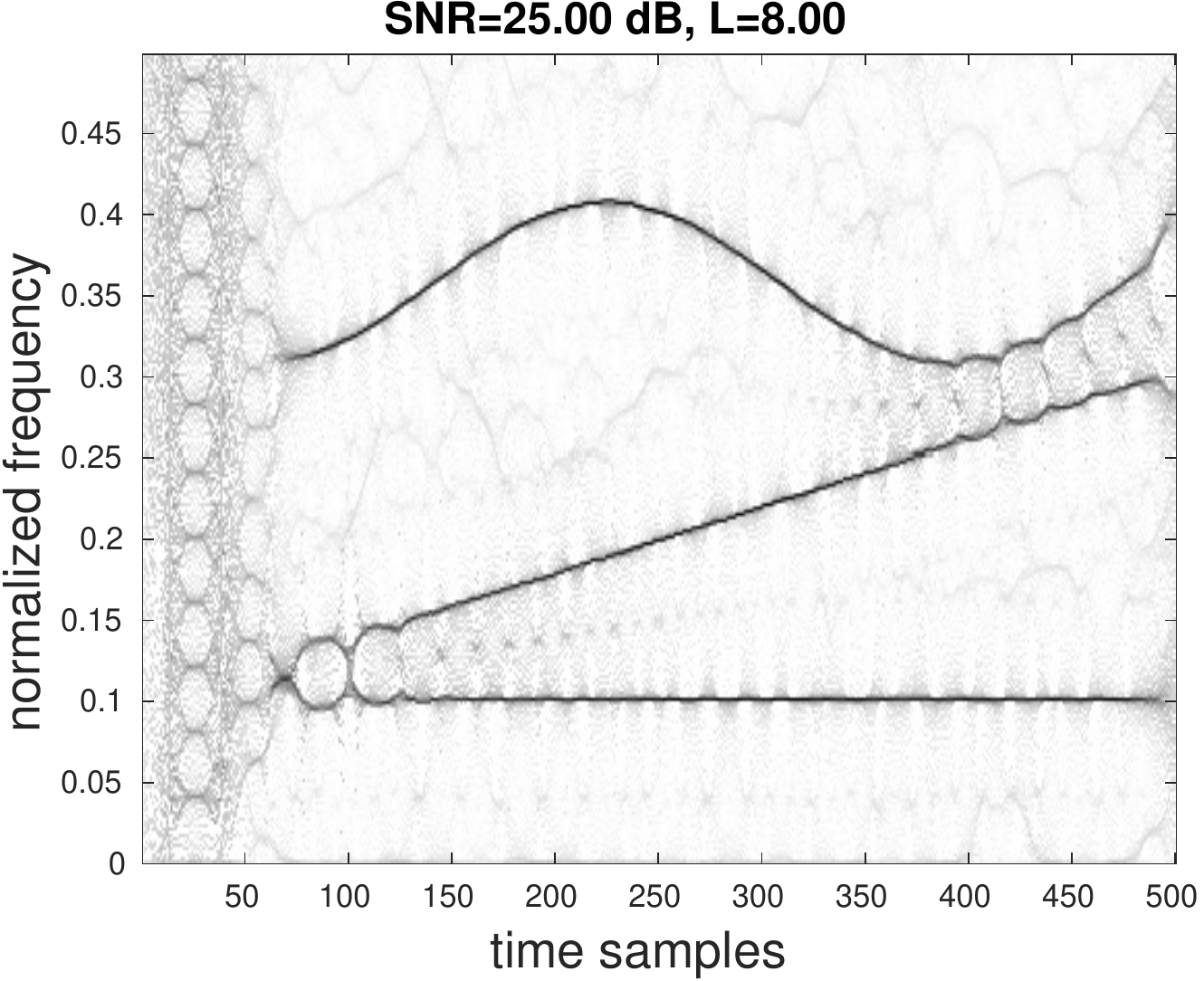}}\\
 \subfigure[reassigned spectrogram]{\includegraphics[width=0.33\textwidth]{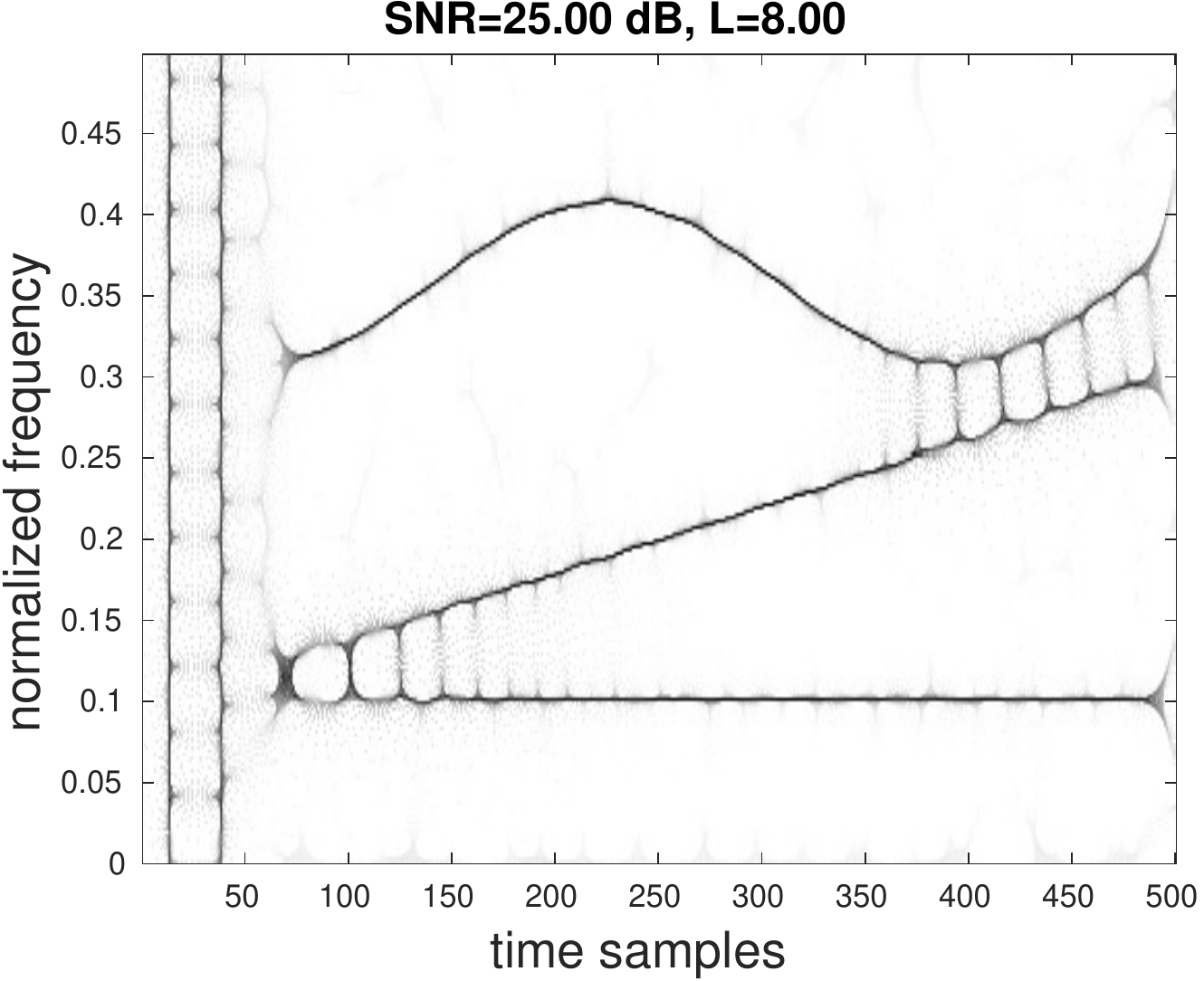}}~\subfigure[time-reassigned synchrosqueezing]{\includegraphics[width=0.33\textwidth]{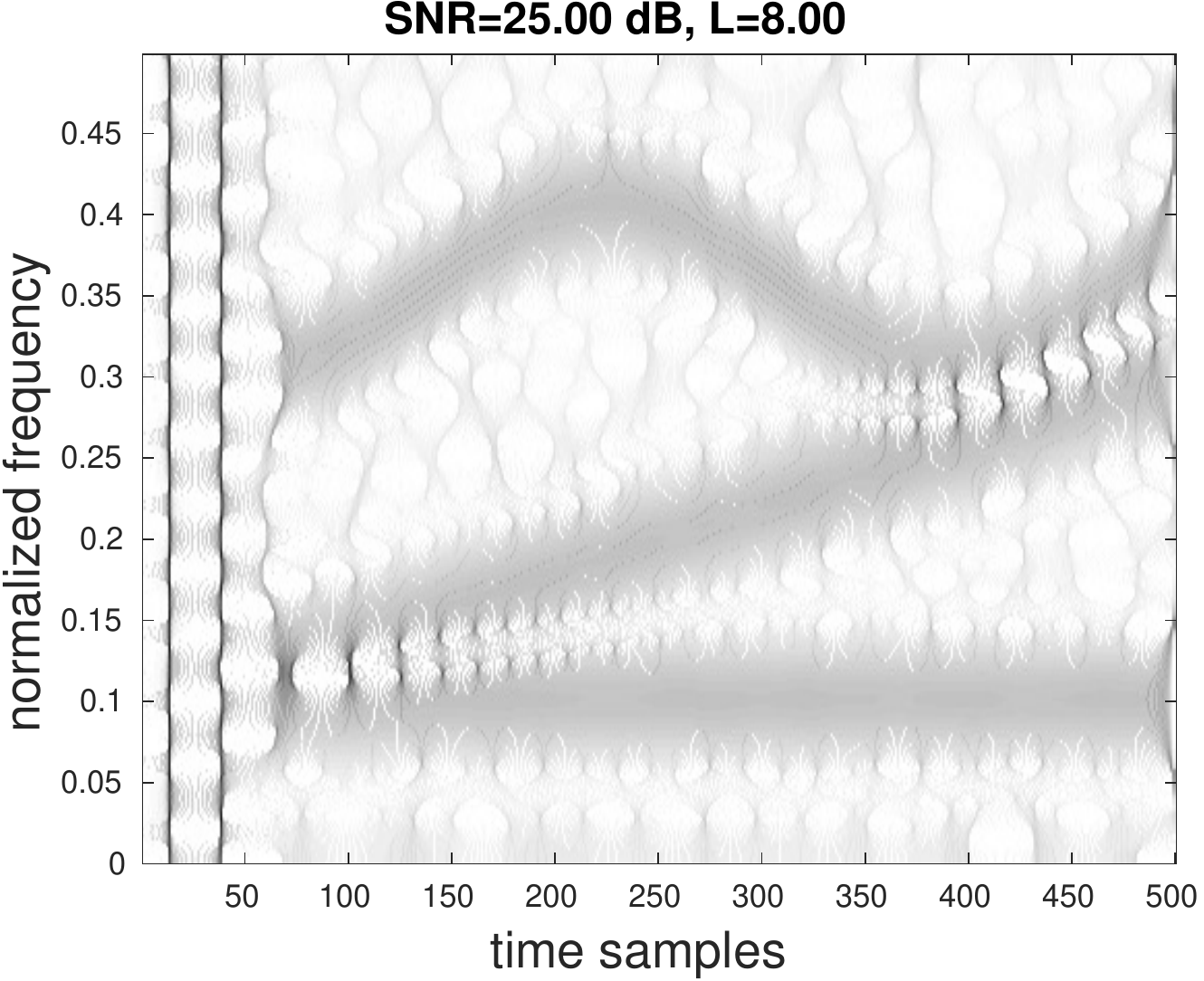}}~\subfigure[second-order horizontal synchrosqueezing]{\includegraphics[width=0.33\textwidth]{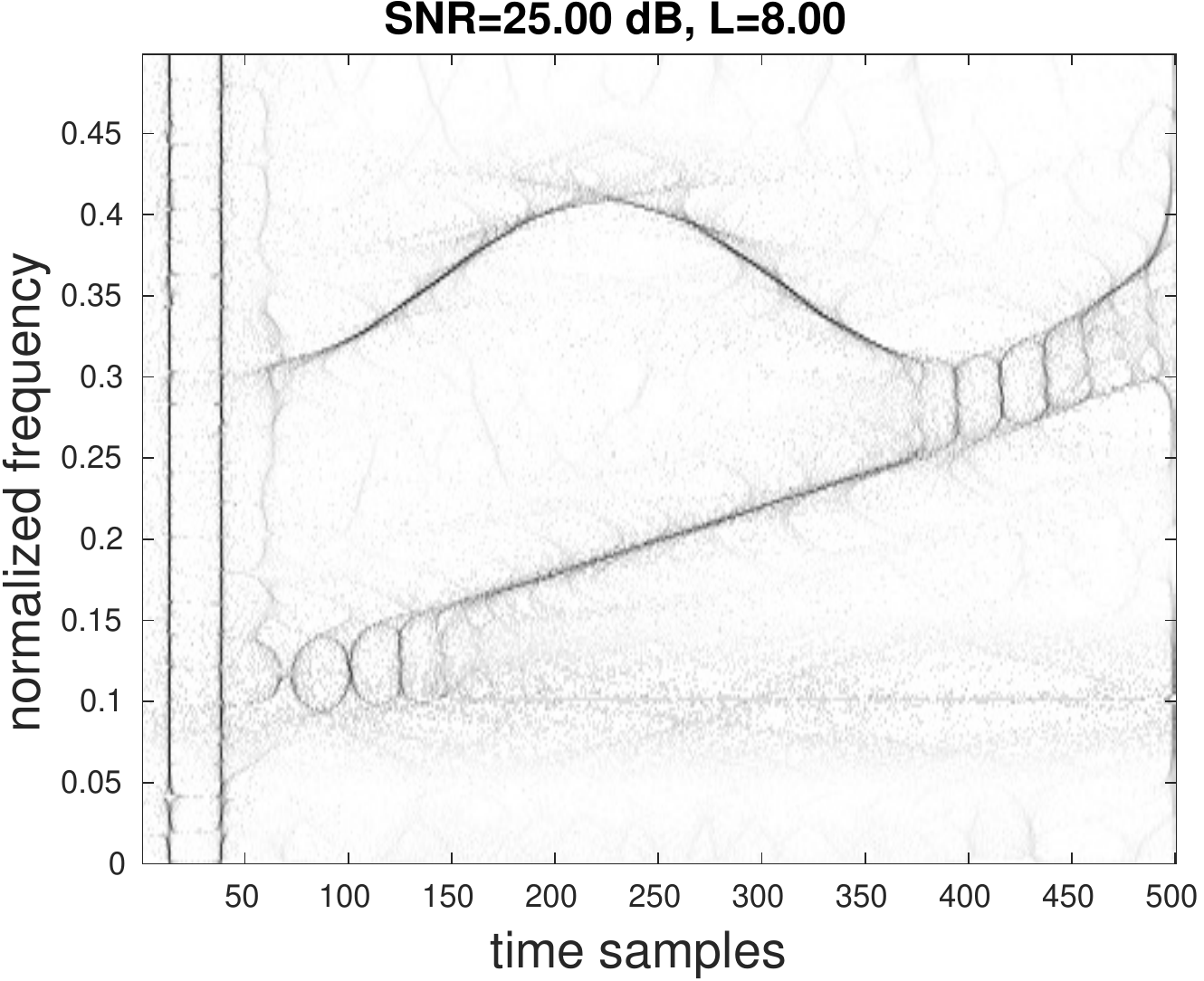}}
\caption{Comparisons of the resulting \ac{tfr}s of a synthetic multicomponent signal. The \ac{tfr}s obtained using the synchrosqueezing methods (b),(c), (e) and (f) correpond to their squared modulus.}
\label{fig:synthetic}
\end{figure*}
%---------------------------------------
\section{Second-order horizontal synchrosqueezing} \label{sec:synchrosqueezing}

\subsection{Enhanced group-delay estimation}

Let's consider a linear chirp signal model expressed as \cite{fourer2017}:
%--------------------------
\begin{align}
		    x(t) 	&= \ee^{\lambda_x(t) + j \phi_x(t)} \label{eq:signal_model}\\
 \text{with}\quad \lambda_x(t)	&= l_x + \mu_x t + \nu_x \frac{t^2}{2}\\
 \text{and}\quad \phi_x(t)	&= \varphi_x + \omega_x t + \alpha_x \frac{t^2}{2}
\end{align}
%--------------------------
where $\lambda_x(t)$ and $\phi_x(t)$ respectively stand for the log-amplitude and phase
and with $q_x = \nu_x + j \alpha_x$ and $p_x = \mu_x + j \omega_x$. For such a signal, it can be shown \cite{fourer2017} that 
$p_x =\tilde{\omega}_x\ttw-q_x\,\tilde{t}_x\ttw$, and therefore:
\begin{equation}
\omega_x = 
\Im(\tilde{\omega}_x\ttw-q_x\,\tilde{t}_x\ttw)
=\hat{\omega}_x\ttw-\Im(q_x\,\tilde{t}_x\ttw)
\end{equation}
The proposed second-order horizontal synchrosqueezing consists in moving $F_x^h\ttw$ from the point $\ttw$ 
to the point ${\scriptstyle(t_x^{(2)},\omega)}$ located on the instantaneous frequency curve, 
\ie such that $\dot{\phi}(t_x^{(2)})\!=\!\frac{d\phi_x}{dt}(t_x^{(2)})=\omega_x + \alpha_x t_x^{(2)}=\omega$. 
This leads to:
%--------------------------
\begin{equation}
t_x^{(2)}\!=\!\frac{\omega-\omega_x}{\alpha_x}=
\hat{t}_x\ttw + \frac{\omega-\hat{\omega}_x\ttw}{\alpha_x} + 
\frac{\nu_x}{\alpha_x}\Im(\tilde{t}_x\ttw)
\label{eq:tx2_fa}
\end{equation}
%--------------------------
which can be estimated by:
\begin{equation}  %\left\{ \begin{array}
\hat{t}^{(2)}_x\ttw\!=\!\! 
\begin{cases} 
\frac{\omega - \hat{\omega}_x\ttw + \Im(\hat{q}_x\ttw\,\tilde{t}_x\ttw)}{\hat{\alpha}_x\ttw}
%\frac{\Im(\hat{q}_x\ttw\tilde{t}_x\ttw) + \omega - \hat{\omega}_x\ttw}{\hat{\alpha}_x\ttw} 
& {\rm if}~\hat{\alpha}_x\ttw\!\neq\!0 \\ 
\hat{t}_x\tw 
& {\rm otherwise}
\end{cases}
\label{eq:t_hat2}% & \text{if} |q_x|>0 \\ \hat{t}_x\tw & \text{otherwise} \end{array} \right.
\end{equation}
%--------------------------
where $\hat{q}_x\ttw=\hat{\nu}_x\ttw + j \hat{\alpha}_x\ttw$ is an unbiased estimator of $q_x$.
This expression can be compared to the second-order group delay estimator introduced by Oberlin 
\etal \cite{second_order_sst}
%--------------------------
\begin{equation}  %\left\{ \begin{array}
\hat{t}^{(2b)}_x\ttw = \begin{cases} \hat{t}_x\tw + \frac{\omega - \hat{\omega}_x\ttw}{\hat{\alpha}_x\ttw} & {\rm if}~\hat{\alpha}_x\ttw\!\neq\!0 \\ \hat{t}_x\tw & {\rm otherwise}\end{cases}. 
\label{eq:t_hat2b}% & \text{if} |q_x|>0 \\ \hat{t}_x\tw & \text{otherwise} \end{array} \right.
\end{equation}
%--------------------------
It can be shown using Eq.\eqref{eq:tx2_fa} \ModifDF{that estimator $\hat{t}^{(2b)}_x$} is biased when $\nu_x\!=\!\frac{d^2\lambda_x}{dt^2}(t) \neq 0$.

Finally, a new second-order horizontal synchrosqueezing transform can thus be obtained
using Eq.\eqref{eq:trsst} by replacing the group-delay estimator $\hat{t}\tw$ by 
our enhanced estimator given by Eq.\eqref{eq:t_hat2}.

\subsection{Theoretical considerations and computation issue}
In \cite{fourer2017,fourer2018b} we introduced two families of unbiased estimators called $(tn)$ and $(\omega n)$
involving $n$-order derivatives ($n\!\geq\!2$) \ModifDF{with} respect to time (resp. to frequency) which enable to compute Eqs. \eqref{eq:t_hat2} and \eqref{eq:t_hat2b}:
%--------------------------------
\begin{align}
 \hat{q}^{(tn)}_x{\tw} 		&= \frac{F_x^{\DD^n h} F_x^h - F_x^{\DD^{n-1} h} F_x^{\DD h} }{F_x^{\TT h} F_x^{\DD^{n-1} h} - F_x^{\TT\DD^{n-1} h} F_x^h } \label{eq:q_x}\\
 \hat{q}_x^{(\omega n)}{\tw} 	&= \frac{(F_x^{\TT^{n-1}\DD h} %{\scriptstyle\tw} 
+ {\scriptstyle(n-1)} F_x^{\TT^{n-2}h}) F_x^h % {\scriptstyle\tw} 
- F_x^{\TT^{n-1} h} %{\scriptstyle\tw}
F_x^{\DD h} %{\scriptstyle\tw}
}
{F_x^{\TT^{n-1} h} %{\scriptstyle\tw}
 F_x^{\TT h} %{\scriptstyle\tw} 
 - F_x^{\TT^n h} %{\scriptstyle\tw} 
 F_x^h %{\scriptstyle\tw}
}\label{eq:crewn}
\end{align}
with $\DD^nh(t) = \frac{d^n h}{dt^n}(t)$ and $\TT^nh(t) = t^nh(t)$. Our preliminary investigations \cite{fourer2018b}
showed a slight improvement using the $(\omega 2)$ estimator in terms of accuracy in comparison to higher-order and $(tn)$ estimators.

Our implementations use the discrete-time reformulations of our previously described expressions 
combined with the rectangle approximation method. Thus $F_x^h[k,m]\!\approx\!F_x^h(\frac{k}{\fs}, 2\pi \frac{m \fs}{M})$, 
where $\fs$ denotes the sampling frequency, $k\in\mathbb{Z}$ is the time sample index and $m \in \mathcal{M}$ is the discrete frequency bin.
The number of frequency bins $M$ is chosen as an even number such as $\mathcal{M}=\left[-M/2+1;M/2\right]$. It results that our method has
the same computational complexity of the previously introduced second-order vertical synchrosqueezing.

The proposed method is valid for any differentiable analysis window. In our implementation\footnote{matlab code freely available at: \url{http://www.fourer.fr/hsst}},
the \ac{stft} uses a Gaussian window and is also called Gabor transform. The window function is 
expressed as $h(t)\!=\!\frac{1}{\sqrt{2\pi}T} \ee^{-\frac{t^2}{2T^2}}$ where $T$ is the time-spread of the window which
can be related to $L=T\fs$.

% EOF

\section{Numerical results} \label{sec:results}

\subsection{Analysis of a synthetic signal}
In this experiment, we consider a synthetic $500$-sample-long multicomponent real-valued signal
made of two impulses, one sinusoid, one chirp and one sinusoidally modulated sinusoid.
Fig.~\ref{fig:synthetic} compares the following \ac{tfr}s: spectrogram, reassigned spectrogram,
classical (frequency-reassigned) synchrosqueezing, second-order vertical synchrosqueezing, time-reassigned synchrosqueezing
and second-order time-reassigned horizontal synchrosqueezing.
Our computations use $M\!=\!600$, $L\!=\!8$ and a \ac{snr} equal to 25 dB obtained by the addition of a Gaussian white noise.
The local modulation estimator $\hat{q}_x^{(\omega 2)}$ is used for computing both second-order synchrosqueezing methods.
The \ac{tfr}s provided by the previously proposed methods are computed using the matlab implementions
provided by the ASTRES toolbox \cite{fourer2017c}.

The results clearly illustrate the improvement of the new second-order time-reassigned synchrosqueezing over the time-reassigned synchrosqueezing
for representing the whole signal.
When compared with frequency-reassigned synchrosqueezing methods, our new method has the advantage to perfectly localize the two impulses
while providing a sharpened representation of the chirp and of the sinusoidally modulated sinusoid. Unfortunately, as for the time-reassigned synchrosqueezing,
our method cannot localize the non-modulated \ModifDF{sinusoid}.% (such signal is rare in the real world).

To assess the signal reconstruction capability, we compare in Table \ref{tab:reconstruction} the \ac{rqf} 
of each \ac{tfr} computed using \cite{fourer2016}: $\text{RQF} = 10 \log_{10}\left( \frac{\sum_n |x[n]|^2}{\sum_n |x[n]-\hat{x}[n]|^2}\right)$.
%--------------------------------------
%\begin{equation}
%\text{RQF} = 10 \log_{10}\left( \frac{\sum_n |x[n]|^2}{\sum_n |x[n]-\hat{x}[n]|^2}\right). 
%\end{equation}
%--------------------------------------
Thus, our results show again the advantage of the time-reassigned synchrosqueezing methods
which obtain significantly higher \ac{rqf} (if $M$ is chosen at least equal to the signal length)
due to its theoretically exact reconstruction formula.
%--------------------------------------
\begin{table}[!ht]
 \caption{Signal reconstruction quality obtained for the reversible \ac{tfr}s presented in Fig.~\ref{fig:synthetic}.}
\centering
\begin{tabular}{|l|c|} 
\hline
%\multirow{2}{*}{Method}					& \multicolumn{2}{|c|}{RQF (dB)}\\
%							& M=500 & M\geq 500 \\
 Method							& RQF (dB)\\
\hline
STFT				 			& 269.27 \\
classical synchrosqueezing 	 			& 35.89  \\
second-order vertical synchrosqueezing 			& 23.80  \\ % 26
time-reassigned synchrosqueezing 			& 116.67 \\
second-order time-reassigned synchrosqueezing   	& 116.67 \\
\hline
\end{tabular}
\label{tab:reconstruction}
\end{table}
%-----------------------------------
\begin{figure*}[!ht]
 \subfigure[signal]{\includegraphics[width=0.33\textwidth]{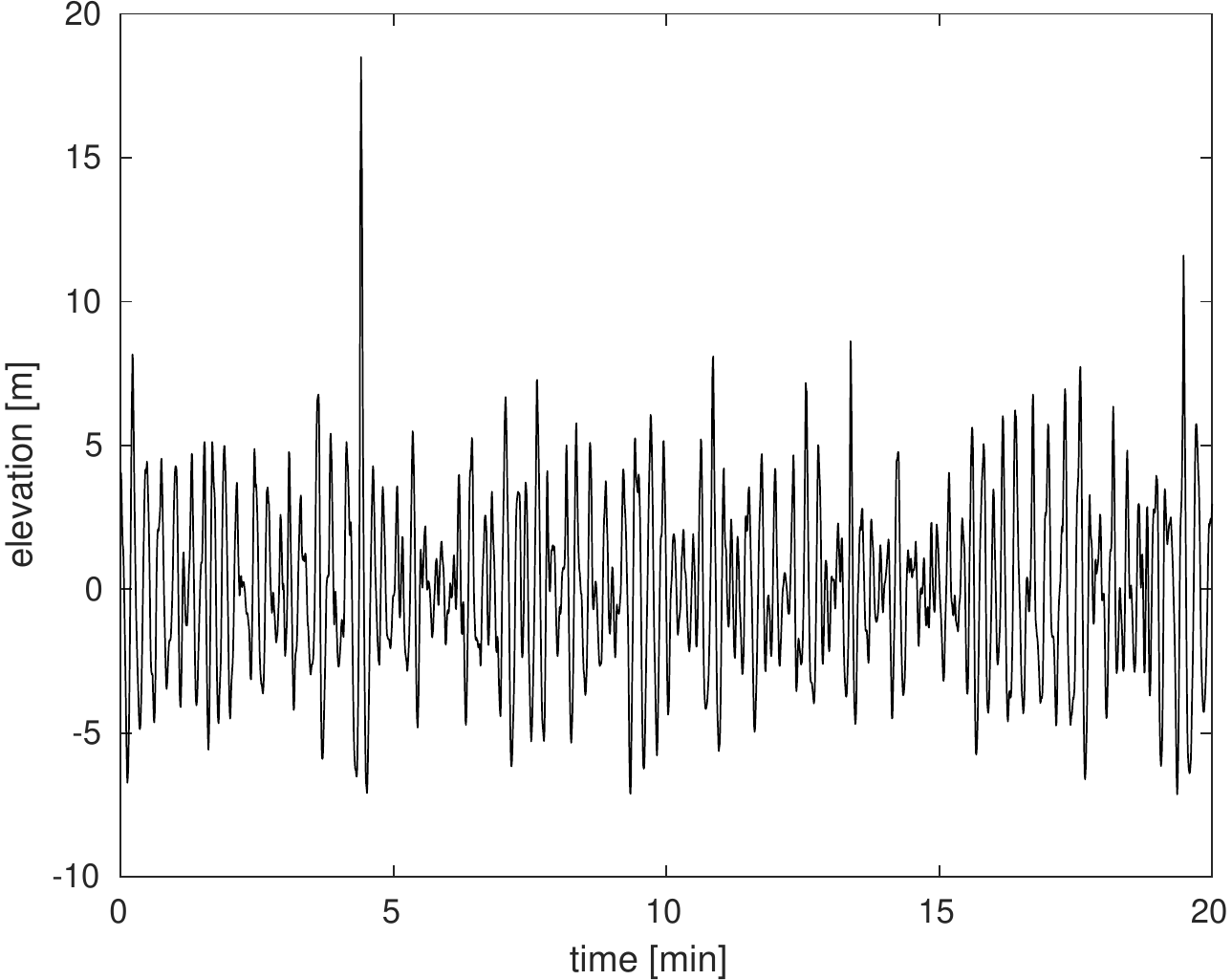}}~\subfigure[synchrosqueezing]{\includegraphics[width=0.33\textwidth]{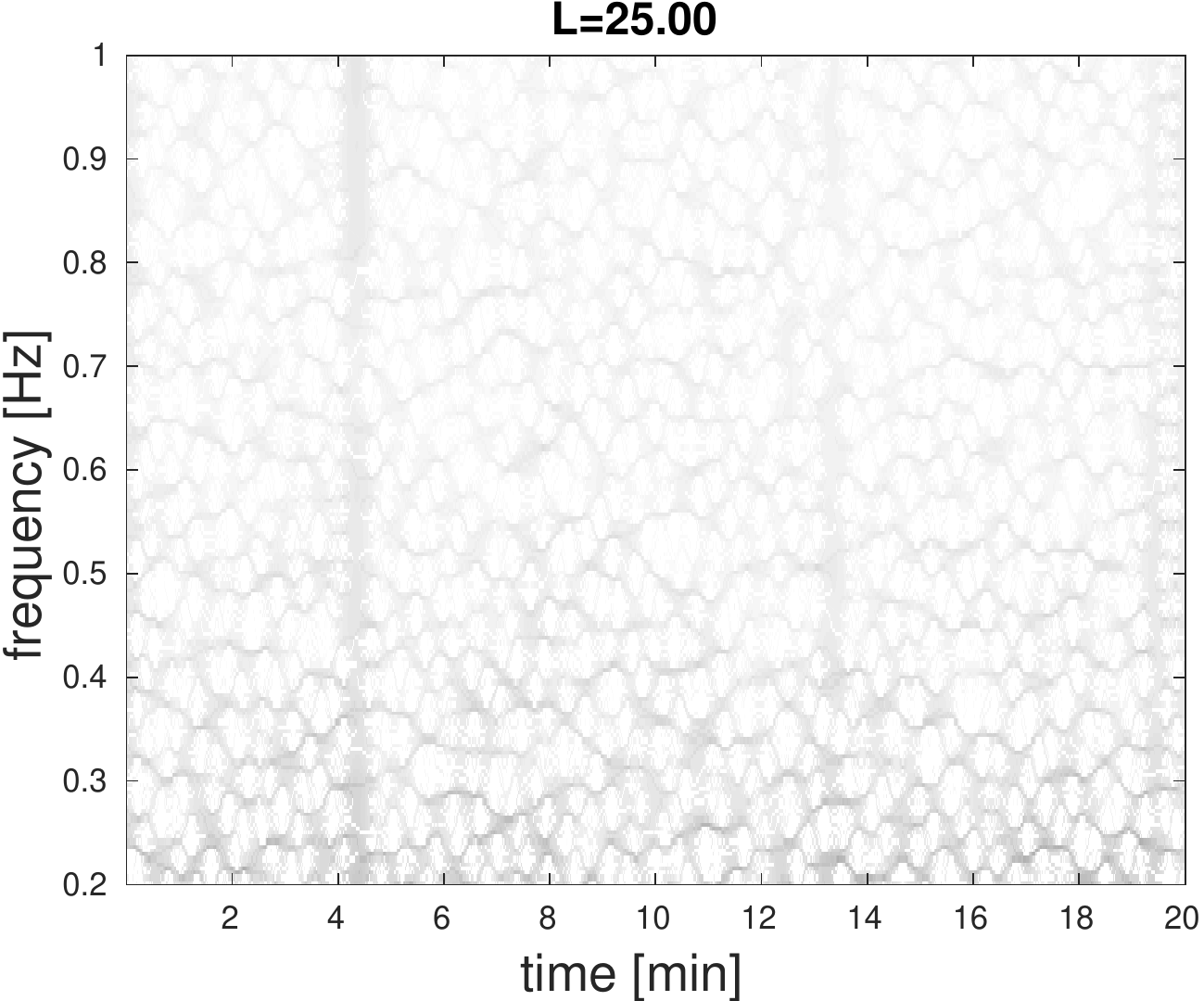}}~\subfigure[second-order vertical synchrosqueezing]{\includegraphics[width=0.33\textwidth]{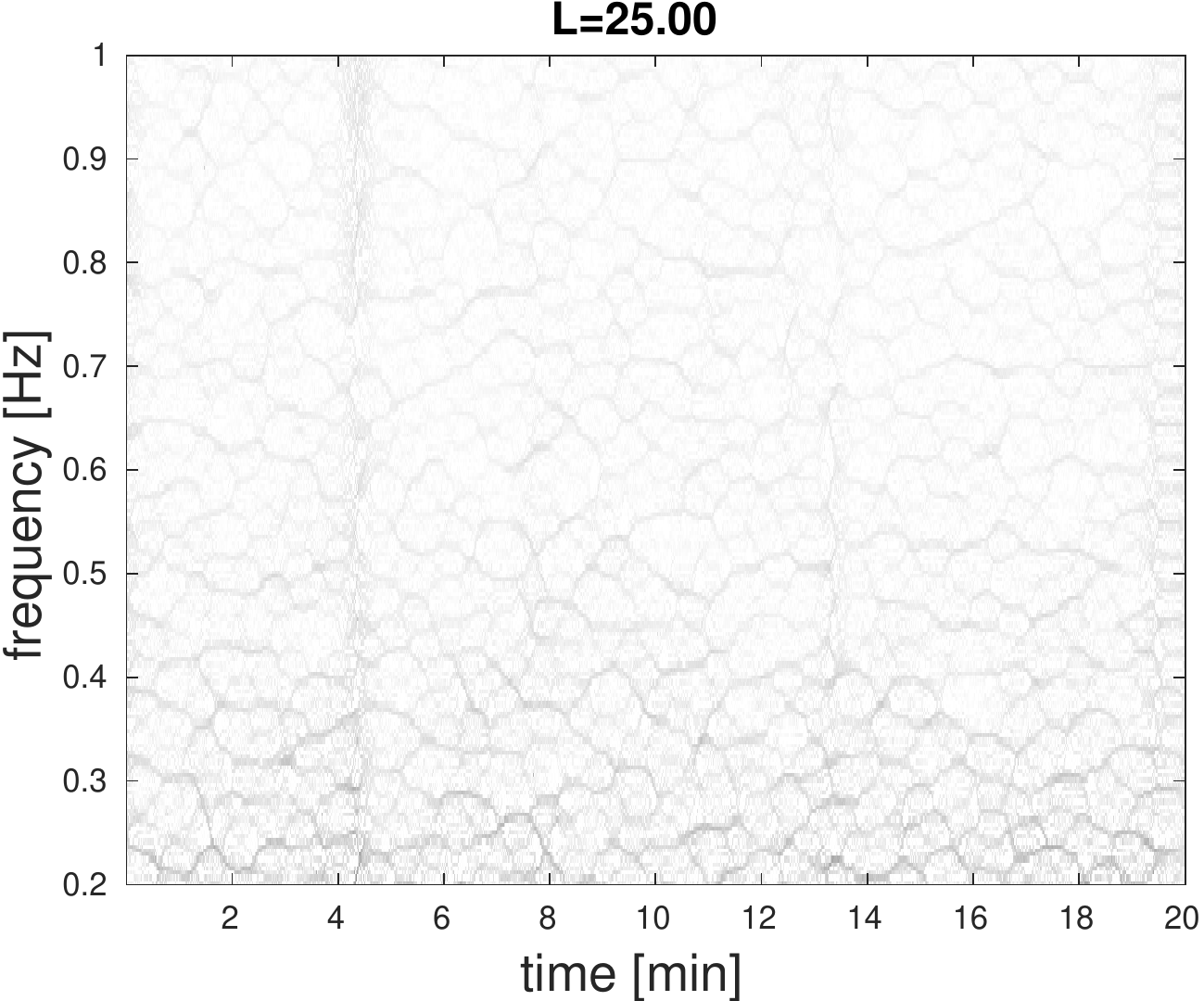}} %~\subfigure[synchrosqueezing]{\includegraphics[width=0.33\textwidth]{figs/synchrosqueezing}}~\subfigure[second-order vertical synchrosqueezing]{\includegraphics[width=0.33\textwidth]{figs/second-order-vertical-SST}}\\
 \subfigure[spectrogram]{\includegraphics[width=0.33\textwidth]{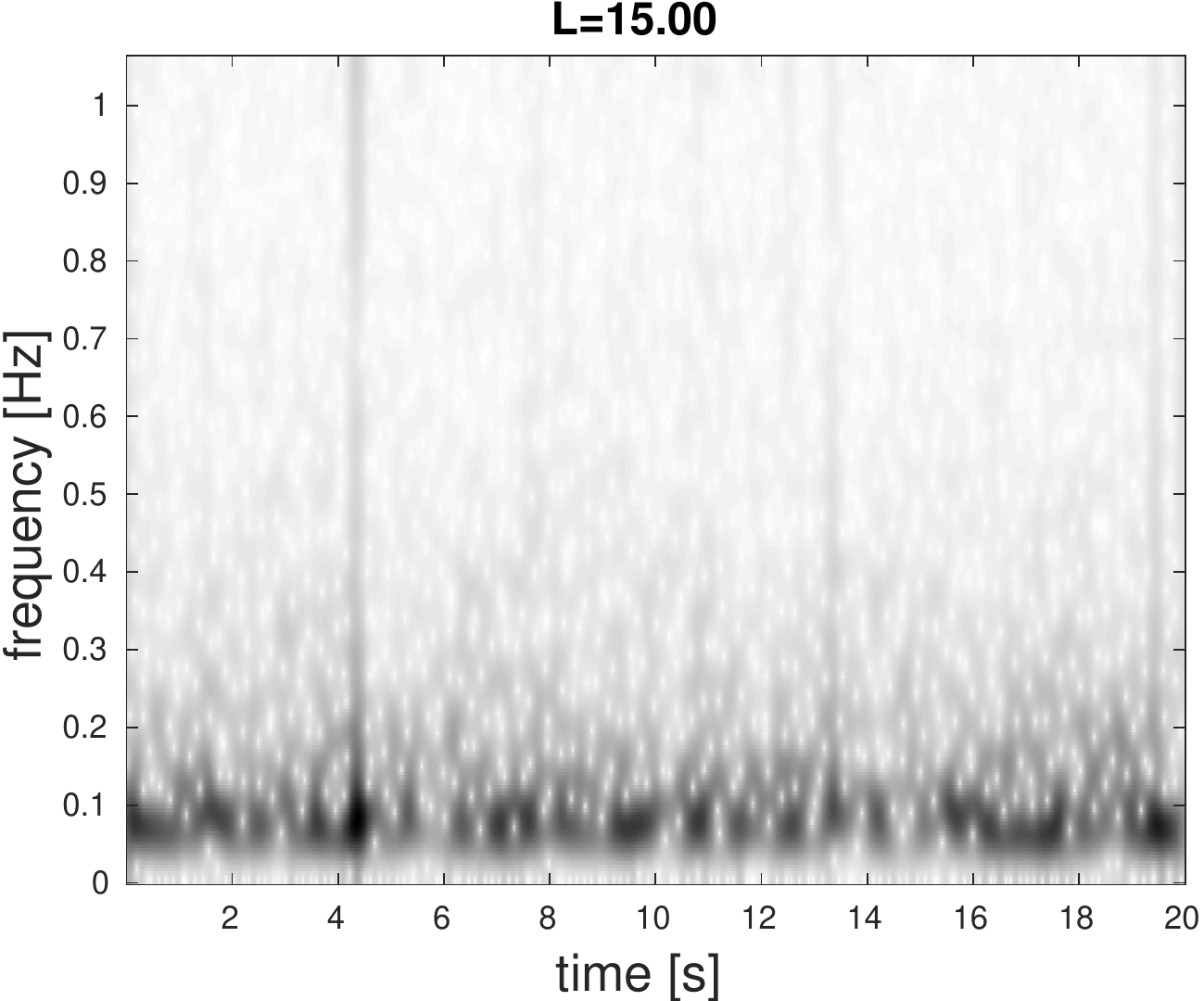}}~\subfigure[time-reassigned synchrosqueezing]{\includegraphics[width=0.33\textwidth]{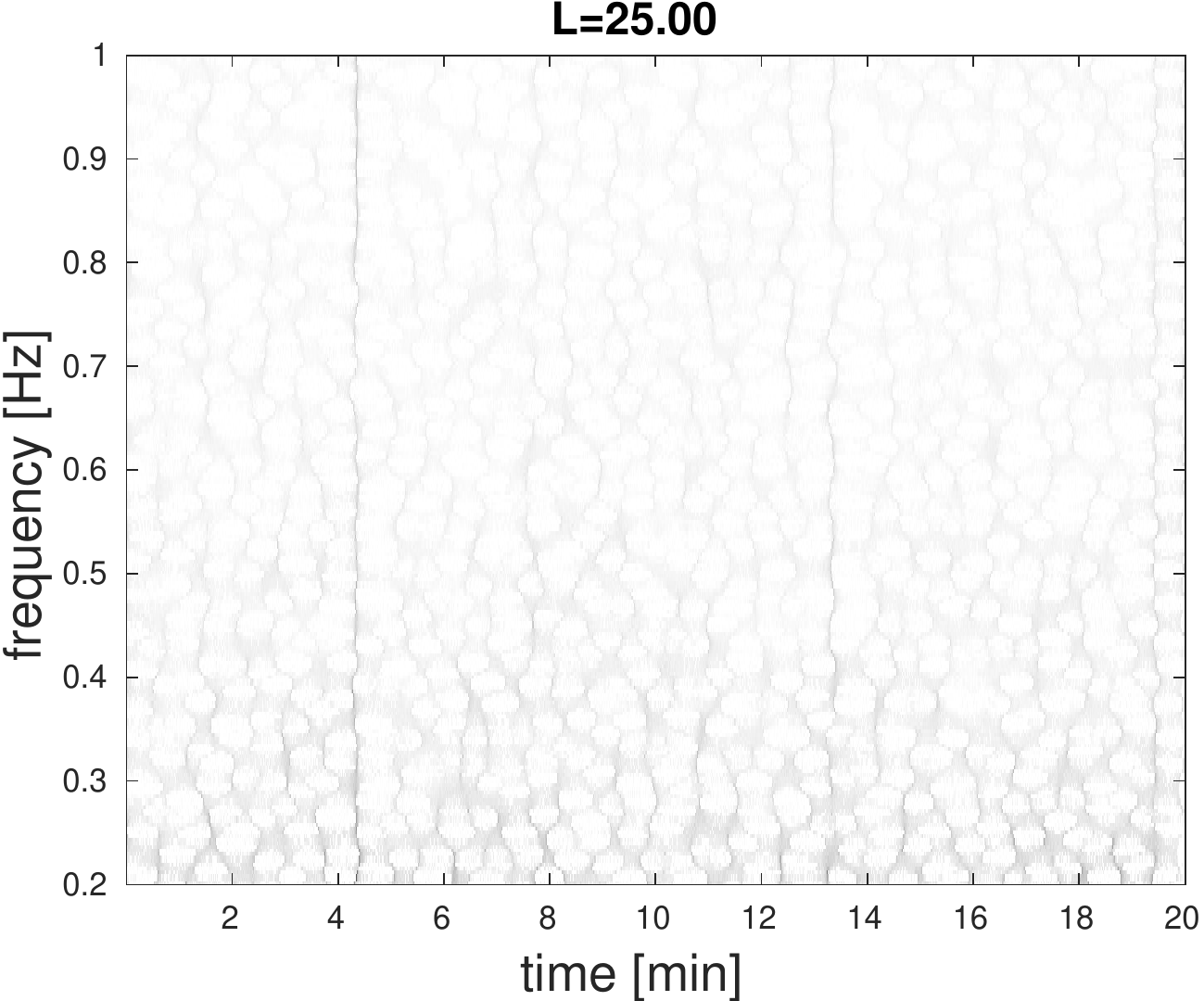}}~\subfigure[second-order horizontal synchrosqueezing]{\includegraphics[width=0.33\textwidth]{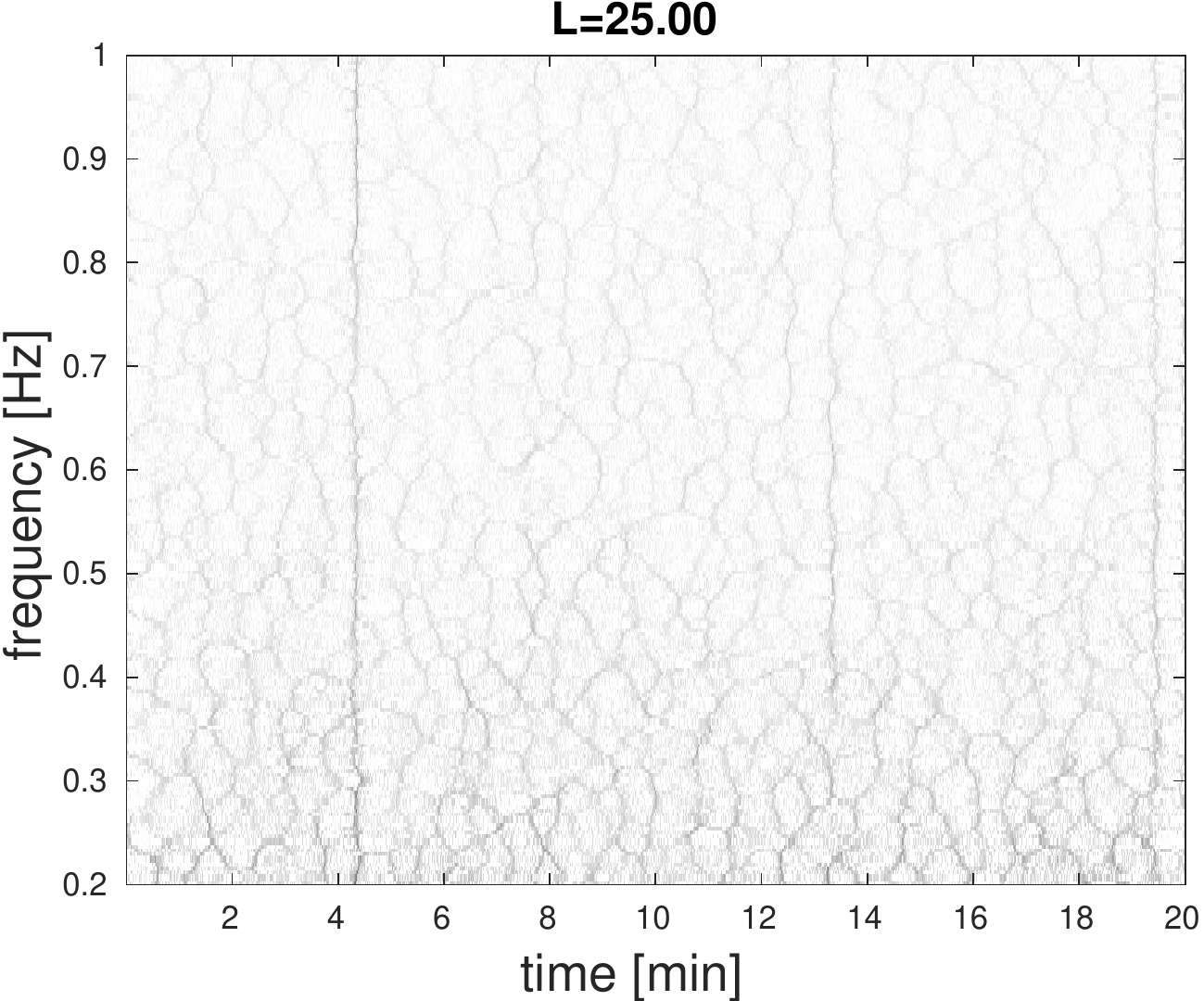}}
\caption{Waveform (a) and \ac{tfr}s of the Draupner wave signal. spectrogram (d), synchrosqueezing (b), second-order vertical synchrosqueezing (c), time-reassigned synchrosqueezing (e) and second-order horizontal synchrosqueezing (f).}
\label{fig:draupner}
\end{figure*}
%-----------------------------------

\subsection{Draupner wave signal analysis}
Now we consider a record of a possible freak wave event measured on the Draupner Platform in 1995 \cite{haver2004possible}. 
The signal displayed in Fig.~\ref{fig:draupner}(a) corresponds to the sea surface elevation deduced from the measures provided 
by a wave sensors consisting of a down-looking laser. The sampling frequency of this signal is $\fs=2.13$~Hz and
its duration is 20 minutes.

\subsubsection{Time-frequency representation}
Fig.~\ref{fig:draupner} compares the resulting \ac{tfr}s provided by the \ac{stft} and its different synchrosqueezed
versions (\ie all the combination of the first- and second-order of the frequency-reassigned and time-reassigned).
For our numerical results, we empirically choose $M=2660$ and $L=25$ which provide sufficiently readable results.
In order to focus to the impulsive part of the signal, we have limited the analysis between 0.2~Hz and 1~Hz.
As expected, the second-order time-reassigned synchrosqueezing provides the best representation
to localize the 4 impulses visible in the signal.
Interestingly, our results reveal the main impulse located at $t_1 \approx 4.39$~min (also visible in Fig.~\ref{fig:draupner}(a))
but also 3 supplementary impulses respectively located at $t_2\approx 7.72$~min, $t_3\approx 13.36$~min and $t_4\approx 19.47$~min.
These impulses were almost invisible in the waveform representation of the signal but have been revealed by our 
proposed time-frequency analysis methods.

\subsubsection{Impulses detection and disentangling}

Now we propose to use the synchrosqueezing signal reconstruction capability for recovering the 4 impulse signals.
To this end, we compute a saliency function defined as the root mean square of the marginal over frequency band  $\Omega=[0.4;1]$~Hz
of the signal energy contained in its synchrosqueezing transform:
%--------------------------------------------
\begin{equation}
 G(t) = \left( \displaystyle\int_{\Omega} |S_x^h\tw|^2 d\omega \right)^{\frac{1}{2}}.
\end{equation}
%--------------------------------------------

\begin{figure}[!ht]
 \centering\subfigure[computed saliency]{\includegraphics[width=0.4\textwidth]{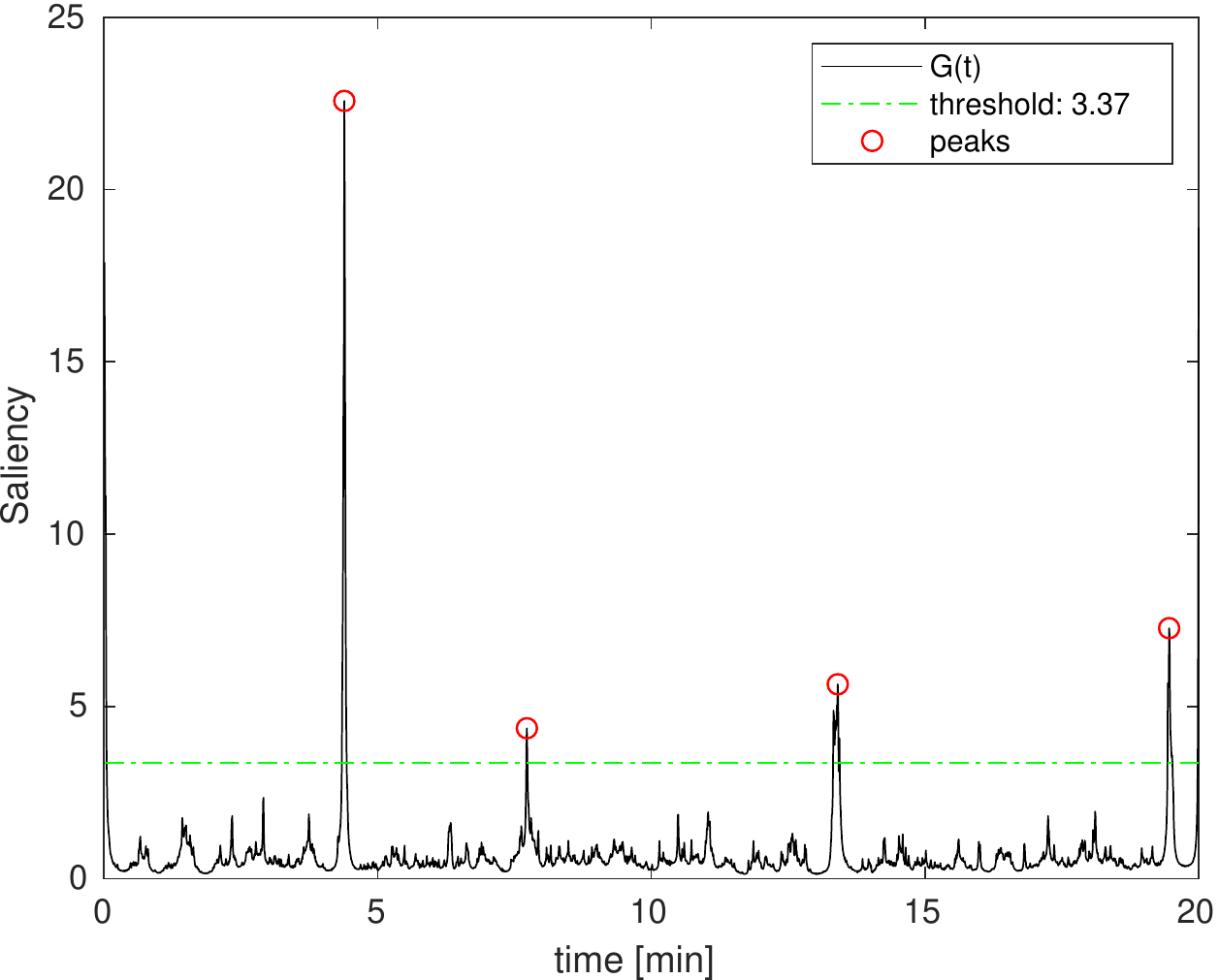}}
 \centering\subfigure[estimated impulses]{\includegraphics[width=0.4\textwidth]{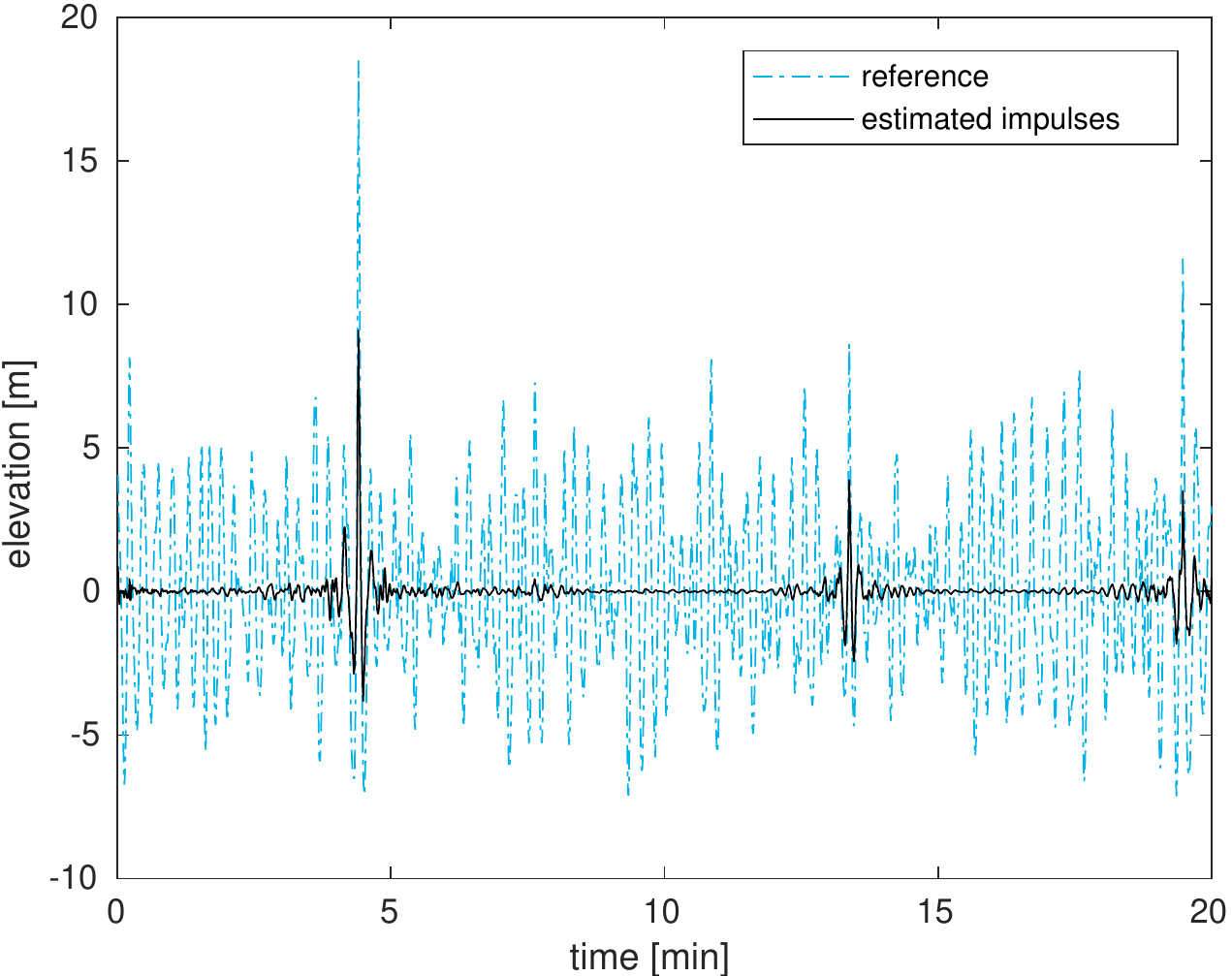}}
\caption{Saliency function $G(t)$ deduced from second-order horizontal synchrosqueezed STFT (a) and reconstructed signal after applying mask on time-reassigned synchrosqueezed STFT.}
\label{fig:detection}
\end{figure}

A binary masked version of the transform $S_x^h\tw$ can thus be computed using $G(t)$ as:
%--------------------------------------------
\begin{equation}
 \hat{S}\tw = \begin{cases} S_x^h\tw & \text{ if } G(t) > \Gamma \\0 & \text{otherwise} \end{cases}.
\end{equation}
%--------------------------------------------
where $\Gamma$ is a defined threshold. Finally the components are extracted by applying
the reconstruction formula given by Eq.\eqref{eq:reconsttsst} on $\hat{S}\tw$.
Our numerical computation presented in Fig.~\ref{fig:detection} uses $\Gamma=3.37$ which corresponds
to 5 times the mean value of $G(t)$. It allows us to recover the impulses locations through a peak picking
and to reconstruct the corresponding waveform signal initially merged in the whole signal.

%% EOF

%---------------------------------------
%---------------------------------------
\section{Conclusion and future work} \label{sec:conclusion}
A new extension of the time-reassigned synchrosqueezing called second-order
horizontal synchrosqueezing was introduced for the \ac{stft}.
Our experiments show a significant improvement to compute invertible and
sharpened time-frequency representations of impulsive signals which cannot be 
addressed by vertical synchrosqueezing.
Moreover, we have shown the efficiency of this technique when applied on both 
synthetic and real-word signals. In fact, our method helped to
discover new signal components in the Draupner wave signal
which could probably help to better understand the phenomenon of freak waves.
Future work consist in theoretically strengthening this method, 
and developing new applications.

%% EOF
\section*{Acknowledgment}
The authors would like to thank Dr. Sverre Haver who kindly agreed to answer our questions and  
share with us the Draupner wave signal record.

%\clearpage

%\nocite{*}



%\bibliographystyle{IEEEtran}
%\vspace{-0.2cm}
\bibliographystyle{IEEEtran}

\end{document}